\documentstyle[12pt]{article}
\advance\textheight by  2.9cm
\advance\topmargin  by -1.5cm
\advance\oddsidemargin by -1cm
\advance\evensidemargin by -1cm
\advance\textwidth by 2.5cm

 \def\z{\mbox{{\cal Z}$[\b A]$}}
 \def\F{\mbox{{\cal F}$[Q,\,\b A]$}}
 \def\kk{\mbox{${\displaystyle \frac{q_\alpha q_\beta}{q^2}}$}}
 \def\G{{\rm g}}
 \def\GG{{\rm {\bf g}}}
 \def\ofi{\mbox{$\overline{\mbox{\boldmath$\varphi$}}$}}
 \def\bfi{\mbox{\boldmath$\varphi$}}
 \def\etab{\mbox{\boldmath$\eta$}}
 \def\s{\mbox{$|\vec s \,\rangle$}}
 \def\S#1{\mbox{$|#1 \rangle$}}
 \def\R{\mbox{$\hat R$}}

\begin{document}

\baselineskip=18pt

\def\thefootnote{\fnsymbol{footnote}}
\begin{center}
 \bf DISTRIBUTIONS OF THE DIFFUSION COEFFICIENT FOR THE QUANTUM AND CLASSICAL
 DIFFUSION IN DISORDERED MEDIA\footnote{Dedicated to Professor
 Hans~A.~Weidenm\"{u}ller on the occasion of his 60th birthday\\
 Nucl.Phys.A560, 274 (1993)} \end{center}

 \begin{center}
 Igor V. Lerner \\ School of Physics and Space Research,\\ University of
 Birmingham, Edgbaston, Birmingham B15 2TT, UK
 \end{center}
 \bibliographystyle{simpl2}
 \def\tr{\mathop{\rm tr}}
 \def\Tr{\mathop{\rm Tr}}
 \def\Im{\mathop{\rm Im}}
 \def\Re{\mathop{\rm Re}}
 \def\ie{{\em i.e.\ }}
 \def\etc{{\em etc}}
 \def\eg{{\em e.g.\ }}
 \def\etal{{\em et al.\ }}
 \newcommand\ef{\varepsilon _{{\!}_F}}
 \newcommand\kf{k_{ { \! }_F}}
 \newcommand\pf{p_{ { \! }_F}}
 \newcommand\nf{n_{ { \! }_F}}
 \newcommand\vf{v_{ { \! }_F}}
 \def\lsim{\raisebox{-.4ex}{\rlap{$\sim$}} \raisebox{.4ex}{$<$}}
 \def\gsim{\raisebox{-.4ex}{\rlap{$\sim$}} \raisebox{.4ex}{$>$}}
 \newcommand\Z{\av{Z[\b h]}}
 \def\bAex{{\bf A}^{ext}}
 \def\Aex{A\r{ext}3}
 \newcommand\r{{\bf{r}}}
 \newcommand\rp{{\bf{r}}^{\!\prime }}
 \def\Ain#1{A^{ext}_{#1}}
  \newif\ifciteref\citereffalse
  \def\citeref#1{{Ref.~\citereftrue\cite{#1}}}
  \def\citerefs#1{{Refs.~\citereftrue\cite{#1}}}
  \def\citenumber#1{{\citereftrue\cite{#1}}}
\def\row#1#2{#1_1,\ldots,#1_{#2}}
\def\rowg#1#2#3{#1\l {1}2#3\ldots #3#1\l {#2}2}
\def\b#1{{\bf #1}}
\def\l#1#2{\lower #2pt\hbox{${\scriptstyle{#1}}$}}
\def\rr#1#2{\raise #2pt\hbox{${\scriptstyle{#1}}$}}
\def\lo#1{\lower 3pt\hbox{${\scriptstyle{#1}}$}}
\def\loo#1{\lower 2.5pt\hbox{${\scriptscriptstyle{#1}}$}}
\def\ra#1#2{\raise #2pt\hbox{$#1$}}

\def\bg#1#2#3{\bigl#1{#3}\bigr#2}
\def\bgg#1#2#3{\biggl#1{#3}\biggr#2}
\def\Bg#1#2#3{\Bigl#1{#3}\Bigr#2}
\def\Bgg#1#2#3{\Biggl#1{#3}\Biggr#2}
\def\mbx#1{\mbox{${\displaystyle #1}$}}
\def\lr#1#2#3{\left#1{#3}\right#2}
\def\bra#1{\left\langle #1\right|}
\def\ket#1{\left| #1\right\rangle}
\def\dbr#1{#1\! #1 }
\def\Dbr#1{#1\!\! #1 }
\def\cum#1{\dbr{\left<}#1\dbr{\right>}}
\def\Cum#1{\Dbr{\Bigl<}#1\Dbr{\Bigr>}}
\def\Cumm#1{\Dbr{\Biggl<}#1\Dbr{\Biggr>}}
\def\cumm#1{\Dbr{\biggl<}#1\Dbr{\biggr>}}
\def\dbp#1{\dbr{\bigl(}#1\dbr{\bigr)}}
\def\av#1{\left <{#1}\right >}
\def\ave#1{\left <{#1}\right >\l {\!\! 0}{12}}
\def\mod#1{{\big |}#1{\bigr |}}
\def\Mod#1{{\Big |}#1{\Bigr |}}
\def\Modd#1{{\Bigg |}#1{\Biggr |}}
\def\modd#1{{\bigg |}#1{\biggr |}}


 \def\sms#1#2{
{\renewcommand{\arraystretch}{0.65}
 \begin{array}{c}
\makebox[0pt]{$\,\,\scriptscriptstyle{#2}$}\\
\makebox[0pt]{$\sum$}\\
\makebox[0pt]{$\scriptscriptstyle{#1}$}
 \end{array}
}\mbox{$\:$}}
 \def\sas#1#2#3{\sum_{
 \begin{array}{c}
\mbox{}\\[-18pt]{\scriptstyle #1}\\[-8pt]{\scriptstyle #2}
 \end{array}
 }^{#3}}

 \def\intr#1#2{{\renewcommand{\arraystretch}{0.4}
 \begin{array}{c}
\mbox{$\;\,\scriptscriptstyle{#2}$}\\
 {\int}\\
\mbox{$\scriptscriptstyle{#1}\;\,$}
 \end{array}
}\!\!\!}
                \newlength{\len}

\def\intg#1#2{{\renewcommand{\arraystretch}{0.4}
                \settowidth\len{$\displaystyle\int$}
                 \begin{array}{c}
                 \makebox[0pt]{\hspace{ \len}$\scriptstyle{#2}$}\\
                 \mbox{${\displaystyle{\int}}$}\\
                 \makebox[0pt]{\hspace{- \len}$\scriptstyle{#1}$}
                 \end{array}
                 }\! }


\def\intp#1#2{{\renewcommand{\arraystretch}{0.4}
                \settowidth\len{$\displaystyle\int$}
                 \begin{array}{c}
                 \makebox[0pt]{\hspace{ \len}$\scriptstyle{#2}$}\\
                 \mbox{${\displaystyle{-\!\!\!\!\!\!\int}}$}\\
                 \makebox[0pt]{\hspace{- \len}$\scriptstyle{#1}$}
                 \end{array}
                 }\! }

                   \newcommand\inte{\intg{0}{\infty}}
                   \newcommand\integ{\intg{-\infty}{\infty}}
                   \newcommand\inter{\intg{1}{\infty }}
                   \newcommand\intpg{\intp{-\infty}{\infty}}
                   \newcommand\intpr{\intp{-1}{1 }}
                   \newcommand\intgr{\intg{-1}{1 }}
                   \newcommand\inp{-\!\!\!\!\!\!\int}

\def\inteb#1{ {\renewcommand{\arraystretch}{0.6}
\begin{array}{c}
 \makebox[0pt]{$\;\;\:\scriptstyle{\infty}$}\\
\mbx{\int}\\
 \makebox[0pt]{$\scriptstyle{#1}\;\;\;$}
 \end{array}
}\!\!\!}

\def\dif#1{\frac{\partial }{\partial #1}}
\def\difa#1#2{\frac{\partial #1}{\partial #2}}
\def\diff#1#2{\frac{\partial^2 }{\partial #1\, \partial #2}}

\def\inteu#1{ {\renewcommand{\arraystretch}{0.6}
\begin{array}{c}
 \makebox[0pt]{$\;\;\:\scriptstyle{#1}$}\\
\mbx{\int}\\
 \makebox[0pt]{$\scriptstyle{0}\;\;\;$}
 \end{array}
}\!\!\!}
                           \def\intbu#1#2{\intg#1#2}



 \newcommand\ir{\int\!\!d^dr}
 \newcommand\ip{\int \!\!\frac{d^dp}{(2\pi )^d}}
 \newcommand\iqd{\int\!\!\frac{d^dq}{(2\pi )^d}}
 \newcommand\ikd{\int\!\!\frac{d^dk}{(2\pi )^d}}
 \newcommand\iq{{(2\pi )^{-2}}\int \!{d^2 q}}

 \def\lmt#1#2#3{{\rm lim}
 {\lower#2pt\hbox{\hspace{#1em}${\scriptstyle{#3}}$}}}


 \def\lim#1#2{{\renewcommand{\arraystretch}{0.65}
 \begin{array}{c}
 \makebox[0pt]{$\,\,$}\\ \mbox{lim}\\
 \makebox[0pt]{$\scriptstyle{#1\rightarrow #2}$}
 \end{array}
 }}

 \begin{quote}
 \small {\bf Abstract.} It is shown that the distribution functions of the
 diffusion coefficient are very similar in the standard model of quantum
 diffusion in a disordered metal and in a model of classical diffusion in a
 disordered medium: in both cases the distribution functions have lognormal
 tails, their part increasing with the increase of the disorder. The
 similarity is based on a similar behaviour of the high-gradient operators
 determining the high-order cumulants. The one-loop renormalization-group
 corrections make the anomalous dimension of the operator that governs the
 $s$-th cumulant proportional to $s(s-1)$ thus overtaking for large $s$ the
 negative normal dimension. As behaviour of the ensemble-averaged diffusion
 coefficient is quite different in these models, it suggests that a possible
 universality in the distribution functions is independent of the behaviour
 of average quantities. %
 \end{quote}

\section{Introduction}
\label{1}

 The absence of self-averaging in mesoscopic systems (see recent
 collections of review pa\-pers \cite{alw,kra}) put forward a question about
 probability distributions of different physical quantities. The conventional
 approach to systems with quenched disorder was based on describing
 measurements in a single sample by the ensemble-averaged quantities, \ie the
 quantities averaged over all realizations of the quenched disorder.
 Discovery of the mesoscopic fluctuations (first of all, of the ``universal''
 conductance fluctuations) manifested that this approach failed for the
 mesoscopic systems, at a length scale quite large as compared to a
 microscopic one. Instead, a statistical approach based on evaluating the
 distribution function of any physical quantity should be implemented. It
 could be reduced to evaluating the average value and the variance only, if a
 distribution is normal. However, a question about the shape of the
 distribution could not be solved {\em a priory}.

 For the problem of quantum diffusion, the mesoscopic scale is defined by
 either the thermal length or the phase-breaking length due to inelastic
 scatterings, both diverging when temperature goes to zero. Then, at $T=0$
 the statistical approach is required for describing any physical quantity at
 any scale. Such an approach, based on the renormalization-group (RG)
 analysis in the framework of the field-theoretical ($\sigma$ model)
 description of the quantum-diffusion problem$^{\mbox{\scriptsize
 \citenumber{Weg:79},\citenumber{EfLKh}}}$), \nocite{Weg:80a} has been
 constructed for different physical quantities at the metallic side of the
 Anderson metal-insulator transition, and extended by means of the
 $2+\epsilon$ expansion (near the critical dimensionality $d=2$) for describing
 these quantities in the vicinity of the transition (see ref.\cite{AKL:91}
 \nocite{IVL:90b} for reviews). A complete description of the transition
 itself should be in terms of changing the distribution of conductance rather
 than of a critical change of the average conductance only, as all
 irreducible moments (\ie cumulants) of the conductance are of the same order
 in the region of the transition so that the conductance distribution is
 completely non-Gaussian. However, in this region the $2+\epsilon$ expansion
 provides quite a poor (at $d=3$) description even for the average
 conductance, as the four-loop calculations by Wegner\cite{Weg:89} proved. In
 view of this, is it possible that the statistical approach based on the
 $2+\epsilon$ expansion could provide at least a qualitative description of
 changing the distribution function in approaching the transition?

 I will argue here that such a possibility does exist. A main qualitative
 prediction of the statistical approach is the appearance of the lognormal
 tails in the distribution functions of different physical quantities, their
 part increasing with the increase of the disorder. For some local quantities
 the distributions become completely lognormal still on the metal side of the
 transition\cite{AKL:91}. It provides a basis for the conjecture that
 statistically the metal-insulator transition reveals itself as a crossover
 from the normal to the lognormal distribution. Such a conjecture seems
 to be confirmed by recent numerical simulations\cite{Spi:90,JC} and
 by the analytical continuation of the one-loop RG results for different
 distribution functions from $d=2+\epsilon$ to $d=1$ that reproduces, after
 substituting a $1$-$d$ value for the mean conductance, exact $1$-$d$ results
 for these functions. It appears, therefore, that although the $2+\epsilon$
 expansion could not provide the reliable values for critical exponents, it
 could describe relations between the cumulants of the same physical
 quantities. The reason is that these relations could be independent of those
 particular properties of the model which determine behaviour of the
 ensemble-averaged quantities (\eg the conductance). An important property of
 the nonlinear $\sigma$ model is the asymptotic freedom. The coupling
 constant (inversely proportional to the conductance) increases making the
 perturbative approach inapplicable in the region of the transition. The
 behaviour of the cumulants, however, is determined essentially by some
 additional operators. For each cumulant, the whole set of the additional
 operators is generated under the RG transformations. The eigenvalues of the
 RG equations that determine the critical exponents for the cumulants are
 governed by the operations of the group of permutations\cite{Weg:90b}
 \nocite{LW:90} acting in this set. These operations seem independent of the
 renormalization of the coupling constant.

 To provide arguments in favour of this viewpoint, I will consider a model of
 classical diffusion in a medium with the quenched disorder. The coupling
 constant of this model is not renormalized to all orders of the loop
 expansion. It does not make the model to be trivial, though, as the
 long-time diffusion is anomalous and characterized by the mean-square
 displacement that increases with time $t$ slower than $t$. Although the
 behaviour of the coupling constant has nothing to do with the asymptotic
 freedom characteristic of the quantum-diffusion problem, the set of the
 eigenvalues of the RG equations governing the cumulants of the diffusion
 coefficients turns out to coincide with that for the quantum-diffusion
 problem. In both cases, the additional operators that govern the cumulants
 are irrelevant within the na\"\i{v}e scaling but become relevant allowing
 for the one-loop RG corrections which give rise to nontrivial deviations of
 the distribution functions from Gaussian. It suggests that such a behaviour
 could be ``universal'' for very different models which by no means belong to
 the same universality class. As the classical-diffusion model is much
 simpler than the quantum one, a further analysis would be possible which is
 quite cumbersome in the quantum case. Due to this similarity, it could shed
 light on some properties of the Anderson transition.

 Before considering the classical-diffusion model, I will describe the
 principal results on the conductance distribution in a disordered metal.

 \section{ Conductance distribution in the quantum diffusion problem}
 \label{2}

 For the sake of comparison with the classical diffusion problem, I will
 outline a way of calculating the moments of conductance distribution using
 the field-theoretical technique. A detailed description can be found in
 ref.\cite{AKL:91}.

 The ensemble-averaged conductance $g$ can be represented directly in terms
 of the generalized nonlinear $\sigma$ model as follows:
 \begin{equation}
 \lr<>{ g_{\alpha,\beta}}=\frac{\pi}{8N^2 L^2}\Tr\biggl\{\frac{\partial^2
 Z[\,\b A]} {\partial \! A_{\alpha}\,\partial \!A_{\beta}}\biggr\}
 \lo{\Bigg|}\l{ {A}=0\atop{N=0} }{12}
 \label{G}
 \end{equation}
 where $L$ is size of the system, $\alpha$ and $\beta$ are the vector indices
 in the $d$-dimensional space, the brackets $\lr<>{\ldots}$ stand for the
 averaging over all realizations of the disorder, and $g$ is chosen to be
 dimensionless quantity measured in the units of $e^2/2\pi^2\hbar$. The
 generating functional is given by %
 \begin{equation}
 \label{ZQ}
 Z[\,\b A ]=\frac{\int\exp\Bigl\{-F\bigl[\,Q;\;\b A\bigr]\Bigl\}{\cal D}Q}
 {\int\exp\Bigl\{-F \bigl[ \,Q\bigr] \Bigl\}{\cal D}Q}
 \end{equation}
 with $F[Q]\equiv F[Q;\b A\! =\! 0]$ being the functional of the standard
 nonlinear $\sigma$ model\cite{Weg:79,EfLKh}
 \begin{equation}
 \label{SM}
 F\bigl[\,Q\bigr]\equiv \frac{1}{\widetilde t}\ir \Tr\,(\partial Q)^2\; .
 \end{equation}
 Here $\partial$ stands for a usual gradient, and the coupling constant
 $\widetilde t\equiv (\pi\nu D/8)^{-1}$ is related to ${\overline{g}\equiv
 \lr<>{g_{xx}}}$ as $\widetilde t=16\pi L^{d-2}/\overline{g}$ where
 $D=\vf l/d$ is the diffusion coefficient, $\nu$ is the one-electron density
 of states, and $l$ is the length of the mean free path. The $2N\times 2N$
 matrix $ Q$ obeys the constraints
 \begin{equation}
 \label{SMC}
 Q^2=I,\qquad\quad\Tr\, Q=0
 \end{equation}
 and the integration in Eq.(\ref{ZQ}) is carried out over independent
 elements of $Q$. In deriving the above field-theoretical representation from
 the initial model of the non-interacting electrons in the random potential,
 the replica trick has been used for the ensemble averaging so that the limit
 $N=O$ should be taken on calculating the conductance (\ref{G}).
 Alternatively, the supersymmetric formulation \cite{Ef:83,VWZ}
 \nocite{Hans:1} could be used for calculating both the average value of
 conductance and its higher moments.

 The functional $F[Q;{\bf A}]$ in Eq.(\ref{ZQ})is obtained from $F[Q]$ by
 substituting the gradient $\partial Q$ by the covariant derivative $\nabla
 Q$:
 \begin{equation}
 \label{CD}
 \partial Q\rightarrow\nabla Q\equiv\partial_\alpha Q-\lr[]{A_\alpha,\,Q},
 \end{equation}
 where the source field ${\bf A}$ is introduced to provide the direct
 representation of the conductance in Eq.(\ref{G}).

 A generalization of Eq.(\ref{G}) for calculating the $s$-th cumulant of the
 conductance is straightforward \cite{AKL:91}:%
 \begin{eqnarray}
 \label{Gs}
 \cumm{\prod_{i=1}^{s}g_{\alpha\l i3\beta\l i3}}=\lr(){\frac{\pi}{8N^2
 L^2}}^{\!s}\biggl\{\prod_{i=1}^{s}\Tr\lr(){\diff{A_{\alpha\l i3}}{A_{\beta\l
 i3}}} \biggr\}\z \lo{\Bigg|}\l{{{ A}=0\atop{N=0}}}{12}.
 \end{eqnarray}
 However, a further extension of the effective field-theoretical model was
 required in Eq.(\ref{Gs}) to obtain all relevant contribution to the
 cumulants. So the generating functional $Z[{\bf A}]$ of Eq.(\ref{ZQ}) has
 been replaced by \z\ obtained by substituting \F\ for $F[Q,\,\b A]$ into
 Eq.(\ref{ZQ}) where the functional \F\ of the extended nonlinear $\sigma$
 model reads %
 \begin{eqnarray}
 \label{ESM}
 \F=F[Q,\,\b A]+\sum_{s=2}^{\infty}F_s[Q,\,\b A]\nonumber \\ F_s[Q,\,\b
 A]=\,z_s \delta\l{\row{\alpha}{2s}}2\ir\Tr\Bigl( \prod_{i=1}^{2n}\nabla\!\l
 {\alpha_i}2 Q\Bigr)\, .
 \end{eqnarray}
 The bare values of the charges $z_s$ are inversely proportional to the bare
 value of the coupling constant $t$. The tensor $\delta\l {\row
 {\alpha}{2s}}2$ in this expression arises from angular integration of the
 product $n\l {\alpha\l 1{2}}2\ldots n\l {\alpha\l {2s}{2}}2$ ($n\equiv {\bf
 p}/\pf$) so that it is proportional to the sum of all the possible products
 of the Kronecker symbols.

 The functional (\ref{ESM}) alongside with the functional of the standard
 $\sigma$ model (\ref{SM}) has been derived microscopically from the initial
 model of free electrons in the random potential $V({\bf r}) $ chosen to be
 Gaussian with zero average and the short-range correlator
 \begin{eqnarray*}
 \bigl< V({\bf r})V({\bf r}^\prime)\bigr>=\frac{1}{2\pi\nu\tau} \delta({\bf
 r}-{\bf r}^\prime)\,
 \end{eqnarray*}
 where $\tau =l/\vf$ is the mean time between elastic scatterings.

 The conductance cumulants (\ref{Gs}) have been calculated with the
 renormalized functional Eqs.(\ref{SM}), (\ref{ESM}). It is important that
 the latter functional is not closed under the RG procedure, and may be
 renormalizable only in a larger space which includes for any integer $s\geq
 2$ (where $2s$ is the total number of $\nabla Q$) all possible vertices of
 the following type:
 \begin{equation}
 F^{\vec {s}}\bigl[\,Q;\;\b A\bigr]= z^{\vec {s}}\,\ir \Bg[]{\Tr\bg(){
 \nabla_{\!\alpha_1\!}Q\nabla_{\!\alpha_2\!}Q}}^{s_1}\ldots \Bg[]{\Tr\bg(){
 \nabla_{\!\beta_1\!}Q\ldots\nabla_{\!\beta_{2m}\!}Q}}^{s_{m}} \ldots
 \label{Fs}
 \end{equation}
 Here a set of integers $ \vec {s} \equiv (\row sm,\ldots) $ obeys the
 constraint
 \begin{eqnarray}
 \sum_{m\ge 1}m s_m\;=s, \label{sm}
 \end{eqnarray}
 and vertices with all possible partitions into the product of traces and all
 possible transpositions of the vector indices $\alpha_i$ and $\beta_i$ (each
 of them being repeated twice) are included. The notation (\ref{Fs}) is
 somewhat symbolic as it means that there are $s_m$ traces of the length $2m$
 while their vector-indices structures are different.

 The bare values of charges attached to all the additional vertices are equal
 zero. However, they are not less ``physical'' then the original ones derived
 microscopically, Eq.(\ref{ESM}). Thus, the vertices which made direct
 contribution to the cumulants (\ref{Gs}) are those with the set $s_1=s$ (and
 $s_m=0$ for $m\neq 1$) rather than the original ones (with the set $s_s=1$
 and $s_m=0$ for $m\neq s$).

 All the high-gradient operators are na\"\i{v}ely irrelevant as the couplings
 $z^{s}\rightarrow \lambda^{d-2s}z^s$ after the scaling $\lambda L\rightarrow
 L$. However, the one-loop RG analysis which reduced, after deriving the RG
 equations for the whole set of the couplings $z^{\vec s}$, to the analysis
 in terms of the group of permutations\cite{AKL:91,Weg:90b} gives the
 dimension $\alpha_s$ of the couplings as follows
 \begin{eqnarray}
 \label{ND}
 \alpha_s=d-2s +\epsilon s(s-1)%
 \end{eqnarray}
 where only the largest eigenvalue $s(s-1)$ of the RG transformations has
 been taken into account. At $d=2$ one substitutes $\epsilon$ by
 $\overline{g}^{-1}$ in this equation.

 There are two different types of contributions to the renormalized values of
 the conductance cumulants. The first one is obtained by the repeated
 differentiation in Eq.(\ref{Gs}) of the generating functional $Z$, which
 includes only the standard functional of the nonlinear $\sigma$ model
 (\ref{SM}) with the substitution (\ref{CD}). The contributions of the second
 type are obtained with allowance for the high gradient vertices
 Eqs.(\ref{ESM}) and (\ref{Fs}). On keeping the most important contribution
 of the second type (which is made for the $s$-th cumulant by the vertices
 with $2s$ gradient operators), one finds \cite{AKL:91} omitting the vector
 indices that
 \begin{eqnarray}
 \label{gn}
 \Cum{g^s}\sim\left\{\
 \begin{array}{llr}
 \overline{g}^{1-s},&s\lsim n_0 \qquad\qquad\qquad &{\rm(a)}\\[-3pt]
 &&\\[-3pt] g_0 (l/L)^{2s-d}\exp\bg[]{u(s^2-s)},&s\gsim n_0&{\rm (b)}
 \end{array}
 \right.
 \end{eqnarray}
 Here $g_0\gg 1$ is the bare value of the conductance,
 \begin{equation}
 \label{u}
 u=\ln\frac{\sigma_0}{\sigma},
 \end{equation}
 $\sigma_0$ and $\sigma$ are the bare and renormalized value of the
 conductivity (not conductance). The number $n_0$ that separates ``high'' and
 ``low'' moments turns out to be quite large in the region of validity of the
 RG approach, $n_0\sim u^{-1} \ln (L/l)$, \ie $n_0\sim g_0$ in the region of
 the weak localization where $\overline{g}=g_0-\ln(L/l)\approx g_0\gg 1$, and
 $n_0\sim g_0/\ln(g_0)$ on approaching the strong localization region where
 $\overline{g}$ drops down to $1$.

 The $\exp(s^2) $-dependence of the moments is a generic feature of the
 lognormal distribution as one can see from the following identity
 \begin{eqnarray}
 \label{LI}
 v^s\exp(us^2)=\frac{1}{(4\pi u)^{1/2}}\inte w^s\,
 \exp\bgg[]{-\frac{1}{4u}\ln^2\frac{w}{v}}\,\frac{dw}{w}.
 \end{eqnarray}
 As only the high moments of the distribution obey this dependence, only its
 tails are lognormal. In the weak-localization region, the low cumulants
 (\ref{gn}a) are small as compared to variance (which is of order 1), and the
 bulk of the distribution is normal. On approaching the strong disorder
 region, the distribution becomes wide and non-Gaussian and the part of the
 lognormal tails increases. However, the small parameter $(l/L)^{2s}$ in
 Eq.(\ref{gn}b) which is due to the normal part of the dimension (\ref{ND})
 of the couplings $z^s$, prevents the tails becoming more essential at the
 metallic side of the transition. In calculating the cumulants of some other
 quantities, like a local density of states (DoS), the negative normal
 dimension of the high-gradient operators is exactly compensated by the
 positive normal dimension of the operators coupled to the cumulants of DoS
 so that the behaviour of the cumulants is governed completely by the
 anomalous dimension. In this case, the distribution function becomes
 completely lognormal when $u\sim 1$ which is still in the metal regime.
 \cite{AKL:91}

 The distribution of the global DoS is similar to that of the
 conductance.\cite{AKL:91} Thus, due to the Einstein relation, the
 distribution of the diffusion coefficient (which, for the quantum diffusion
 problem, has not been directly calculated) has similar features. I will show
 below that the same behaviour of the distribution of the diffusion
 coefficient is characteristic of some model for classical diffusion in
 disordered media.

 \section{ Classical diffusion in media with the quenched disorder}
 \label{3}

 Diffusion of classical particles in random media has attracted considerable
 attention (see ref.\cite{CDif} for a recent review).
 \nocite{Sin,Dan1,ArNel,Dan2,KLY:85,KLY:86b} A set of
 models$^{\mbox{\scriptsize \citenumber{Sin}--\citenumber{KLY:85}}}$) is
 given by the Langevin equation that describes random walks of a particle
 under the influence of a thermal noise $\etab(t)$ in the medium with the
 quenched disorder governed by a random-drift field ${\bf v}(\r)$:
 \begin{equation}
 \label{L}
 \dot{\r} = {\bf v}(\r) +\etab(t)%
 \end{equation}
 The thermal noise $\etab$ is chosen to be Gaussian white noise with a zero
 average and
 \begin{equation}
 \label{WN}
 \overline{\eta_\alpha(t)\,\eta_\beta(t')}=2D\,\delta_{\alpha\beta}\,
 \delta(t-t'),
 \end{equation}
 where $D$ is the diffusion coefficient.

 Equivalently, averaging over the random noise one describes these random
 walks in terms of the Fokker-Planck (FP) equation
 \begin{equation}
 \label{FP}
 \lr[]{\dif t +\partial_\alpha\lr(){v_\alpha-D\partial_\alpha}}\rho(\r,t)=0
 \end{equation}
 with $\rho(\r,t)$ being the local density of the diffusing particles.

 The exact solution obtained by Sinai \cite{Sin} to the model
 Eqs.(\ref{L})--(\ref{FP}) in one dimension (with the Gaussian short-range
 correlator for the random-drift fields $v$) has revealed the anomalous
 diffusion manifested by the strongly sub-diffusive long-time behaviour of the
 random walks. Later such a diffusion has been considered as a possible
 universal source for the excess ($1/f$) current noise \cite{MPRW}.

 However, it appeared that anomalous diffusion occurs only for $d\leq 2$, at
 least in the weak-disorder case, as $d=2$ turned out to be the upper
 critical dimensionality, and the long-time behaviour of the random walks
 could be super-diffusive as well as sub-diffusive, depending on some
 features of the model. The point is that a vector character of the quenched
 disorder in $d>1$ gives more freedom in choosing correlator of the fields
 ${\bf v}$. In general, assuming the Gaussian nature of the random-drift
 fields, \ie irrelevancy of the higher-order correlations (the assumption
 that could be proved in a subsequent RG analysis), and choosing $\lr<>{\b
 v}=0$, one may write has
 \begin{equation}
 \label{VC}
 \lr<>{v_\alpha(\r)v_\beta(r')}=\gamma_0F_{\alpha\beta}(\r-\r')
 \end{equation}
 where $\gamma_0$ characterizes the strength of the disorder.

 The simplest possible choice of the correlator \cite{Dan1}
 \begin{equation}
 F_{\alpha\beta} =\delta_{\alpha\beta}\delta(\r-\r')
 \label{F1}
 \end{equation}
 leads to the model where corrections to the effective diffusion coefficient
 $D(t)$ vanish in the long-time limit. Two other specific models for
 $F_{\alpha\beta}$ arise naturally$^{\mbox{\scriptsize
 \citenumber{ArNel}--\citenumber{KLY:85}}}$) on imposing constraints on the
 random field $\b v$ choosing it either potential ($v_\alpha=\partial_\alpha
 \Phi$) or solenoidal ($ \partial_\alpha v_\alpha(\r)=0$). For these models
 the Fourier transform of the correlator (\ref{VC}) reads
 \begin{eqnarray}
 \qquad F_{\alpha\beta}=\left\{\begin{array}{rll} \kk, &\mbox{potential
 field}\qquad\qquad\qquad &{\rm (a)} \\[-3pt] &&\\[-3pt]
 \delta_{\alpha\beta}-\kk, &\mbox{solenoidal field}\qquad\qquad\qquad &{\rm
 (b)}\end{array}\right.
 \label{Fk}
 \end{eqnarray}
 The RG analysis has shown$^{\mbox{\scriptsize
 \citenumber{ArNel}--\citenumber{KLY:85}}}$) that at $d=2$ the potential
 disorder leads to the sub-diffusive behaviour in the long-time limit while
 the solenoidal disorder leads to the super-diffusive one:
 \begin{eqnarray}
 D(t)\equiv \dif t \lr<>{\r^2(t)}\propto \left\{ \begin{array}{lll}
 D_0(t/\tau)^{-\GG_0/2}, &\mbox{potential field} \qquad\qquad\qquad
 &(a)\\[-3pt] &&\\[-3pt] D_0\ln^{1/2}(t/\tau), &\mbox{solenoidal field}
 \qquad\qquad\qquad &(b)\end{array}\right.
 \label{As}
 \end{eqnarray}
 Here
 \begin{equation}
 \label{g}
 \G_0=\frac{\gamma_0}{4\pi D_0^2}
 \end{equation}
 is the bare value of the coupling constant $\G$\ (dimensionless at $d=2$)
 which plays a role of the effective weak-disorder parameter, $D_0$ is the
 bare value of the diffusion coefficient (describing diffusion at $t\sim
 \tau$), and $\tau$ is some small time which provides the ultraviolet cutoff
 necessary to regularize the model.

 Note that the results of Eq.(\ref{As}) are asymptotically exact in the
 weak-disorder limit. In the case of the solenoidal disorder, the coupling
 $\G$ is decreased by the RG transformation from its bare value $\G_0\ll 1$
 (the ``zero-charge'' situation) which makes the strong-disorder region
 unattainable. In the case of the potential disorder, there is no
 renormalization of the coupling constant $\G$ at all. It has been firstly
 demonstrated up to the two-loop order of the RG analysis \cite{KLY:85} but
 has later been proved \cite{KLY:86c} to be perturbatively exact: the
 renormalization is absent in all the orders of the loop expansion. It does
 not make the model trivial as both the strength of the disorder $\gamma$ and
 the diffusion coefficient $D$ (as well as other physical quantities, such as
 mobility) are renormalized which leads, in particular, to the sub-diffusion,
 Eq.(\ref{As}a).

 The quenched disorder may enter the FP equation (\ref{FP}) also via the
 spatial variation of the diffusion coefficient $D(\r)\equiv \overline{D}
 +2\delta D(r)$ where $\overline{D}$ is the uniform part of $D(\r)$. One can
 take the limit $\b v=0$ and define the disorder in terms of the distribution
 $P(D)$ of the spatially uncorrelated diffusion coefficients. The
 distribution is governed by its cumulants
 \begin{eqnarray}
 \Cum{\delta{D}(\r_1)\delta{D}(\r_2)}&=&2!\,g^{(2)}\delta(\r_1-\r_2),
 \nonumber\\ \cdots&&\cdots\label{DCu}\\
 \Cum{\delta{D}(\r_1)\ldots\delta{D}(\r_s)}&=&s!\,g^{(s)}\delta(\r_1-\r_2)
 \ldots \delta(\r_1-\r_s),\nonumber
 \end{eqnarray}
 where the additional couplings $g^{(s)}$ could have changed under the
 renormalization.

 Such a model was believed to be trivial in the weak-disorder limit as the
 operators described the quenched fluctuations of the diffusion coefficient
 are irrelevant under the RG transformations \cite{KLY:85} at any
 dimensionality $d>0$ so that the long-time diffusive behaviour remains
 normal and the limiting distribution of the diffusion coefficients turns out
 to be Gaussian whatever is the initial choice of the constants $g^{(s)}$.
 However, as it has been recently suggested by Tsai and Shapir \cite{Ysh},
 this model could be analyzed in terms of the $d=0+\epsilon$ expansion which
 could shed some light at the diffusive behaviour in the strong-disorder
 limit. In the case of the annealed disorder, the analysis of ref.\cite{Ysh}
has
 shown that in the ``strong-disorder'' limit (\ie for some nonzero
 $\epsilon$) a nontrivial fixed distribution of the diffusion coefficient
 existed which turned out to be the inverse Gaussian one. Unfortunately, it
 appeared to be impossible\cite{Ysh} to have extended the results of this
 analysis to the more interesting case of the quenched disorder that
 corresponds to the $N=0$ replica limit. (The annealed limit corresponds to
 the $N=1$ case, \ie to the averaging of the non-replicated functional).

 Notwithstanding a viability of $0+\epsilon$ expansion for describing the
 strong disorder in any physical dimensionality and the lack of analysis in
 the case of the quenched disorder, the appearance of the nontrivial fixed
 point \cite{Ysh} seems to be quite important. However, the
 ``zero-dimensional'' model defined by the FP equation (\ref{FP}) with ${\bf
 v}=0$ and by the distribution (\ref{DCu}) has some quite specific features.
 The most important is that all the operators coupled to the diffusion
 cumulants become marginal at the critical dimensionality $d=0$ (\ie the
 {na}\"\i{ve} scaling dimension for each operator becomes equal zero).
 In any model with $d\neq 0$, the higher the order of cumulant, the more
 irrelevant becomes the appropriate operator. Thus, in the model of the
 quantum diffusion described in the Section (\ref{2}) the {na}\"\i{ve}
 dimension of the $2s$-th high-gradient operator (which defines the
 $s$-th cumulant of the conductance and, to some extent, of the diffusion
 coefficient) decreases proportionally to $-s$ as in Eq.(\ref{ND}). It makes
 impossible an impact of the operators with the higher number of gradients on
 those with the lower one that, in turn, makes implausible the appearance of
 a new fixed point for the distribution (at least, starting from the
 weak-disorder limit where the dimensional analysis has some sense). It does
 not mean, however, that the distribution remains trivial (\ie normal). As
 was shown in the Section (\ref{2}), to the one-loop order the dimension
 of the high-gradient operators becomes proportional to $s^2$ so that they
 could become relevant. It leads to the increasing (\ref{gn}) of the higher
 cumulants which define nontrivial tails of the distribution. As these tails
 become more essential on approaching the strong disorder regime, the
 shape of the distribution could have changed completely in the
 strong-disorder limit \cite{AKL:91}. As all the cumulants have different
 critical exponents (which are not proportional to the order of the
 cumulant), the change in the shape of the distribution has a character of a
 smooth crossover that is not associated with a new fixed point.

 The goal of further considerations is to show that a similar scenario can
 take place in a classical-diffusion model. I will consider the FP equation
 (\ref{FP}) that includes both the random drifts ${\bf v}(r)$ defined with
 the correlator Eqs.(\ref{VC})--(\ref{Fk}), and the random diffusion
 coefficients defined with the cumulants (\ref{DCu}).

 Before proceeding with the renormalization, I will describe a possible
 realization of the random-drifts models.

 \subsection{ RELATION TO LATTICE HOPPING MODELS}

 The lattice hopping model which goes over to the FP equation (\ref{FP}) in
 the continuum limit is defined by the master equation for the probability
 $\rho_\r(t)$ of finding the particle at the $\r$-th site at the moment $t$:
 \begin{equation}
 \label{ME}
 \difa{\rho_\r}{t}=\sum_{\r'}\bg(){W_{\r\r'}\rho_{\r'}-W_{\r'\r}\rho_\r}. %
 \end{equation}
 Here $W_{\r\r'}$ is the probability of hopping from site $\r'$ to site $\r$
 which slightly fluctuates around the hopping probability $W^0$ in a regular
 lattice:
 \begin{equation}
 \label{W}
 W_{\r\r'} =W^0_{\r-\r'}+\delta{W}_{\r\r'},%
 \end{equation}
 with $\delta{W_{\r\r'} }$ describing the quenched weak disorder on the
 lattice.

 The parameters of the continuum model described by the FP equation
 (\ref{FP}) are readily expressed in terms of the hopping probabilities:
 \begin{eqnarray}
 \label{DW}
 D(\r)&=&\frac{1}{2} \sum_{\r'}(\r-\r')^2W_{\r\r'}\\
{\bf v}(\r)&=& \quad\sum_{\r'}(\r-\r')W_{\r\r'}
\label{vW}
\end{eqnarray}
 The random drift velocity given by Eq.(\ref{vW}) vanishes in all
 realizations of the disorder until the asymmetry in the hopping
 (\mbox{$\delta W_{\r\r'}\neq \delta W_{\r'\r}$}) is included. Therefore, the
 symmetric hopping model goes over to the FP equations with ${\bf v}=0$ where
 the disorder in $D$ described by the cumulants (\ref{DCu}) reflects the
 symmetric disorder in $W$. In this case, the analysis of the continuum limit
 showed the weak disorder to be irrelevant for any $d$.

 The asymmetry could be due to the presence of magnetic or charged impurities
 which leads to the continuum model with either the solenoidal or the
 potential disorder. Following ref.\cite{KLY:85}, I will show
 that in the presence of a random electric field ${\bf E}(\r)$ of charged
 impurities the continuum-limit model includes the potential random drifts
 (\ref{Fk}a). If ${\bf E}$ is the only source of randomness, the
 probability of thermally activated hops at a distance $b$ is
 \begin{equation}
 \label{hop}
 W_{\r, \r+\b b} = W_\b b ^0 \exp \lr[]{-\frac{e\b E(\r)\b b}{2kT}}.
 \end{equation}
 Assuming then a random distribution of the charged impurities on the lattice
 and a global electro-neutrality, and representing the Fourier transform of
 ${\bf E}(\r)$ as
 $$
 {\bf{E}}({\bf q})=\frac{2\pi i{\bf q}}{\varepsilon q} \sum_{j}e_j\exp(i{\bf
 q\cdot r}_j),
 $$
 with $e_j=\pm e$ being the charges of impurities with random coordinates $\b
 r_j$ and $\varepsilon$ being the dielectric constant, one finds
 \begin{eqnarray}
 \lr<>{E_\alpha({\bf q})E_\beta({\b q'})}=\frac{(2\pi e)^2C}{\varepsilon}\kk
 \delta_{\bf q q'}
 \label{Ecu}
 \end{eqnarray}
 where $C$ is the concentration of the charged impurities. In the continuum
 limit, using the linear in $\b E$ approximation ($\delta W\propto E$) and
 assuming for simplicity the nearest-neighbor hopping only, one arrives at
 the potential-random-drifts model with the correlator Eqs.(\ref{VC}),
 (\ref{Fk}). The strength of disorder $\gamma_0$ in Eq.(\ref{VC}) is related
 to the parameters of the hopping model as
 \begin{equation}
 \label{g0}
 \gamma_0=\frac{\pi^2 e^4 C}{\varepsilon^2(kT)^2}\lr(){\frac{W_0z a^2}{2}}^2
 \equiv \frac{\pi^2 e^4 C}{\varepsilon^2(kT)^2} D_0^2%
 \end{equation}
 where $z$ is the coordination number of the lattice, $a$ is the lattice
 constant, $W_0\equiv W_a^0$ is a regular part of the nearest-neighbor
 hopping probability, and $D_0$ is the bare value of the diffusion
 coefficient. The randomness in $W$ results also in some distribution $P(D)$
 of the diffusion coefficient whose cumulants (\ref{DCu}) could be directly
 expressed in terms of the cumulants of $\delta W$.

 In real lattices the correlator (\ref{Ecu}) becomes non-singular ($\propto
 q_\alpha q_\beta$) for $q\gg r_0$ where $r_0$ is a screening radius. The
 continuum model with the correlator (\ref{Fk}a) could then describe only the
 random walks at a distance not exceeding $r_0$. Some models where $r_0$ is
 macroscopically large are described in ref.\cite{KLY:85}. In general, $r_0$
 takes the part of the infrared cutoff similar to that of the phase-breaking
 length in the weak-localization theory and defines the mesoscopic scale for
 the classical-diffusion problem. Below I will discuss the continuum model
 with the unscreened correlator (\ref{Fk}).

 \subsection{ THE EFFECTIVE FUNCTIONAL}

 Random walks of the particle started at the initial moment $t=0$ at the
 point $\r$ is characterized by the probability to find it at the point $\r'$
 in time $t$%
 \begin{equation}
 \label{tG}
 {G(\r',\r;t)}=\lr<>{\rho(\r',t)\,\rho(\r,0)}%
 \end{equation}
 which is the Green's function of the FP equation (\ref{FP}). The asymptotic
 properties are governed by the long-time behaviour of this function, or by
 low-frequency behaviour of its Fourier transform, $G(\r,\r';\omega)$. To
 perform the ensemble averaging, it is convenient to represent the latter as
 a functional integral over the conjugate complex fields
 $\overline{\varphi}(\r)$ and $\varphi(\r)$:
 \begin{equation}
 \label{FI}
 G(\r,\r';\omega)=\frac{i\int \overline{\varphi}(\r)\varphi(\r')
 e^{iS[\overline{\varphi},\varphi]}{\cal D}\overline{\varphi}\, {\cal
 D}\varphi} {\int{e^{iS[\overline{\varphi},\varphi]} {\cal D}
 \overline{\varphi}\, {\cal D}\varphi}},%
 \end{equation}
 where the effective action functional is given by
 \begin{equation}
 \label{A2}
 S[\overline{\varphi},\varphi] = \ir \bg[]{i\overline{\varphi}\omega\varphi
 +v_\alpha\bg(){\partial_\alpha \overline{\varphi}}\varphi -D\partial
 _\alpha\overline{\varphi} \partial _\alpha\varphi}%
 \end{equation}

 The averaging is performed by the standard replica trick: the fields
 $\overline{\varphi}(\r)$ and $\varphi(\r)$ and the functional integration in
 Eq.(\ref{FI}) are $N$-replicated and the independent averaging over the
 numerator and denominator in Eq.(\ref{FI}) is justified in the replica limit
 $N=0$ (that should be taken in the final results). Taking into account the
 disorder both in the random drifts ${\bf v}$ and in the diffusion
 coefficients $D$ which is defined by Eqs. (\ref{VC}) and (\ref{DCu}), one
 deduces the effective action
 \begin{equation}
 \label{EA}
 {\cal S}[\ofi,\bfi] =\Bg\{\}{{\cal S}_0 +{\cal S}_{int} +{\cal
 S}_{cum}}[\ofi,\bfi] %
 \end{equation}
 which should be substituted for that given by Eq.(\ref{A2}) into
 Eq.(\ref{FI}) where the functional integration should be performed over all
 components of the fields $\bfi\equiv \row\varphi{N}$ and $\ofi\equiv
 \row{\overline{\varphi}}{N}$. Here
 \begin{eqnarray}
 \label{S0}
 {\cal S}_0[\ofi,\bfi] &=&\ir\,
 \ofi\bg(){i\omega+\overline{D}\partial^2}\bfi\\%
 \label{Si}
 {\cal{S}}_{int}[\ofi,\bfi]&=&\frac{i\gamma}{2}\ir d^d r'\,
 \bg(){\partial_\alpha\ofi\,\bfi}_{\r}
 F_{\alpha\beta}(\r-\r')\bg(){\partial_\beta\ofi\bfi}_{\r'}\\%
 \label{Sa}
 {\cal S}_{cum}[\ofi,\bfi]&=&i\sum_{s=2}^{\infty}g^{(s)}{\cal S}_{(s)};\qquad
 {\cal S}_{(s)}= \ir\,\prod_{i=1}^{s}
 \bg(){\partial_{\alpha_i}\ofi{\partial}_{\alpha_i}\bfi}_\r
 \end{eqnarray}
 The RG analysis of the functional (\ref{S0}), (\ref{Si}) in the upper
 critical dimensionality $d=2$ has been used for describing the anomalous
 long-time behaviour of the {\em average} diffusion coefficient,
 Eq.(\ref{As}).

 The higher-order gradient operators, Eq.(\ref{Sa}), describe the
 renormalization of the cumulants (\ref{DCu}) of the distribution $P(D)$.
 Under rescaling $L\rightarrow \lambda{L}$, $g^{(s)}\rightarrow
 \lambda^{(s-1)d}g^{(s)}$ so that the {na}\"\i{ve} dimension of $g^{(s)}$
 behaves like that for the high-gradient operators in the quantum diffusion
 problem, and $g^{(s)}$ are irrelevant. I will show now that in the case of
 the potential random drifts, Eq.(\ref{Fk}a), the one-loop RG corrections
 overturn this conclusion and make the scaling dimensions of the high-order
 gradient operators positive.

 \subsection{ ONE-LOOP RENORMALIZATION}

 The renormalization is performed by expanding exp$(i{\cal S}_{int} +i{\cal
 S}_{cum})$ in a power series, integrating with the weight $\exp\bg(){i{\cal
 S}_0[\ofi_0,\bfi_0]}$ over the ``fast'' components of the fields (this
 integration will be denoted as $\lr<>{\ldots}_0$), and exponentiating the
 results of the integration. Here $\bfi(\r)$ (and $\ofi(\r)$) is decomposed
 into the sum of ``slow'', ${\widetilde{\bfi}}(\r)$, and ``fast'',
 $\bfi_0(\r)$ components where
 $$
 {\widetilde{\bfi}}(\r) =\sum_{q<\lambda q_0}\bfi({\bf q})e^{i{\bf
 q}\cdot\r}, \qquad \bfi_0(\r) =\sum_{\lambda q_0<q<q_0}\bfi({\bf q})e^{i{\bf
 q}\cdot\r},%
 $$
 $0<\lambda<1$ is the scaling parameter, $q_0$ is the ultraviolet cutoff, \eg
 the inverse lattice constant in the lattice realization of the model. In the
 one-loop order (\ie in the first order in powers of the coupling constant
 $\G$, Eq.(\ref{g})), the logarithmic (at $d=2$) contribution to $g^{(s)}$
 is made only by the term
\begin{equation}
 -\lr<>{{\cal S}_{int}{\cal S}_{(s)}}_0\equiv
 - \G\Lambda{\hat{R}}\Bg\{\}{{\cal S}_{(s)}}. %
 \label{av}
 \end{equation}
 Here the one-loop RG operator $\hat R$ is introduced, $\Lambda\equiv
 \ln\lambda^{-1}$. The relevant contractions (\ie those which give the
 logarithmic term) have the structure
 \begin{equation}
 \lr<>{\varphi_0^a\partial_\alpha\overline{\varphi}^{\,b}_0}_0
 \label{ff}
 \lr<>{\varphi_0^c\partial_\beta\overline{\varphi}^{\,d}_0}_0,
 \end{equation}
 where fields $\varphi_0$ are taken from ${\cal S}_{int}$, and fields
 $\overline{\varphi}_0$ from ${\cal S}_{(s)}$. As $\lr<>{\overline{\varphi_0}
 \varphi_0}_0$ is just a diffusion propagator, the contractions containing
 more gradients are irrelevant. That is why there is no contributions from
 the terms like
 $
 \lr<>{{\cal S}_{cum}^n}_0,
 $
 and no feedback from the terms with the higher number of gradients to those
 with the lower one.

 It is evident from Eq.(\ref{ff}) that the index structure of the action
 (\ref{Sa}) is not conserved under the renormalization. A set of additional
 operators is generated having the structure
 \begin{eqnarray}
 \partial_\alpha\overline{\varphi}\,^a\partial_\beta\varphi^b
 \partial_\gamma\overline{\varphi}\,^c\partial_\delta\varphi^d\ldots
 \end{eqnarray}
 with all possible permutations of the vector indices $\alpha,\beta,\ldots$
 and the replica indices $a,b,\ldots$, each index being repeated
 twice, where the summation over repeated replica indices from $1$ to $N$ and
 over repeated vector indices from $1$ to $d$ is implied. These operators are
 ``unphysical'' in a sense that the distribution $P(D)$ is defined by the
 renormalization of the initial operators (\ref{Sa}) only. Then, only those
 operators are of interest which have the RG feedback to the initial ones.

 To classify all the additional operators one introduces matrix notations
 \begin{equation}
 \label{MN}
 {\cal Q}^{ab}=
 \partial_\alpha\overline{\varphi}\,^a\partial_\alpha{\varphi}^b,\qquad
 {\cal{P}}^{ab}= \partial_\alpha{\varphi}^a\partial_\alpha{\varphi}^b,\qquad
 \overline{{\cal P}}\,^{ab}=
 \partial_\alpha\overline{\varphi}\,^a\partial_\alpha\overline{\varphi}\,^b.
 \end{equation}
 In these notations, the contractions (\ref{ff}) could involve only the
 matrices ${\cal Q}$ and $\overline{{\cal P}}$. On calculating the action of
 the RG operator (\ref{av}) on the product ${\cal QQ}$ one finds
 \begin{equation}
 \label{RG}
 {\hat{R}}\Bg\{\}{{\cal Q}^{ab}{\cal Q}^{cd}} = \Bg[]{ A( {\cal{Q}}^{ab}{\cal
 Q}^{cd} +{\cal Q}^{ad}{\cal Q}^{cb}) +(2-|A|)\overline{\cal P}\,^{ac}{\cal
 P}^{bd}}%
 \end{equation}
 Here $A$ discriminates the models described above: $A=1$ for the potential
 random drifts, Eq.(\ref{Fk}a), $A=-1$ for the solenoidal ones,
 Eq.(\ref{Fk}b), and $A=0$ for the spatially uncorrelatted random drifts,
 Eq.(\ref{F1}).

 As $\hat R$ does not act on the matrices ${\cal P}$ (they do not contain the
 field $\overline{\varphi}$ to be contracted with the field $\varphi$ from
 the ${\cal S}_{int}$), their number in any operator could not be reduced
 under the renormalization. Therefore, the RG equations have a triangular
 structure: the operators containing only the matrices ${\cal Q}$ are not
 influenced on by those containing $\overline{{\cal P}}{\cal P}$ (the number
 of $\overline{{\cal P}}$ is evidently equal to the number of ${\cal P}$ in
 any operator).

 The initial operator (\ref{Sa}) is proportional to $\lr[]{\Tr{\cal Q}}^s$,
 so that one can restrict the RG analysis to the subset of the operators
 containing only the matrices ${\cal Q}$. There is no one-loop corrections to
 $g^{(s)}$ in the case of the uncorrelated random drifts, Eq.(\ref{F1}), as
 only the $\overline{{\cal P}}{\cal P}$ term arises on renormalizing the
 initial action (\ref{Sa}). I consider the case of the potential random
 drifts, Eq.(\ref{Fk}a). The action (\ref{RG}) of the RG operator $\hat R$ on
 the chosen subset reduces to simple permutations of the matrix indices of
 ${\cal Q}$. It can be analyzed in terms of the group of permutations,
 similar to the case of quantum diffusion \cite{AKL:91,Weg:90b}. Any operator
 containing $s$ matrices ${\cal Q}$ (\ie $2s$ gradients of the fields
 $\overline{\varphi}$ and $\varphi$) may be written down in a form similar to
 that for $2s$-gradient operators in the quantum-diffusion problem (compare
 to Eq.(\ref{Fs})):
 \begin{eqnarray}
 {\cal S}_{\vec s}[{\cal Q}] = g^{(\vec s)}\ir\lr\{\}{\Bg[]{
 \Tr\bg(){\cal{Q}}}^{s_1}\ldots\Bg[]{\Tr\bg(){\cal{Q}}^{m}}^{s_m}\ldots}
 \label{QQ}
 \end{eqnarray}
 Here the set of integers $\vec s$ obeys the constraint (\ref{sm}), the
 initial operator (\ref{Sa}) corresponding to $\vec s=(0,\ldots, 0,s)$. The
 renormalization of the $s$-th cumulants results from solving the RG
 equations for the whole set of $g^{(\vec s)}$, the bare values of all the
 additional charges being equal to zero.

 The RG equations may be written down as follows
 \begin{equation}
 \label{RGE}
 \frac{d\s}{d\Lambda} = \G\R\s%
 \end{equation}
 where $g^{(\vec s)}$ is represented as some ket-vector defined by the
 ``occupation numbers'' $s_m$
 \begin{equation}
 \label{s}
 g^{(\vec s)}\equiv\s\equiv\S{1^{s_1}\,2^{s_2}\ldots{m^{s_m}}\ldots}, %
 \end{equation}
 and $\hat R$ is expressed in terms of the ``creation'' and ``annihilation''
 Bose operators $a^{\dagger}$, $a$
 \begin{eqnarray}
 \lefteqn{\R = \frac{1}{2} \biggl\{\Bg[]{\sum_{m\ge
 1}m\,a_m^{\dagger}a_m}^2-\sum_{m\ge 1}m\,a_m^{\dagger}a_m}\nonumber \\ &&+
 \sum_{m,l\ge{1}}\Bg[]{(m+l)\,a_m^{\dagger}a_l^{\dagger}a_{m+l}+ml\,
 a_{m+l}^{\dagger}a_ma_l}\biggr\},
 \label{R}
 \end{eqnarray}
 these operators acting on the states \s\ as follows:%
 \begin{eqnarray}
 \label{aa}
 \begin{array}{lcll} a_m\S{1^{s_1}\ldots{m^{s_m}}\ldots} &=&\!\!s_m\!\!\!\!&
 \S{1^{s_1}\,\ldots{m^{s_m-1}}\ldots}\\[-3pt]
 &&&\\[-3pt]a_m^{\dagger}\S{1^{s_1}\ldots{m^{s_m}}\ldots}
 &=&&\S{1^{s_1}\ldots{m^{s_m+1}}\ldots} \end{array}
 \end{eqnarray}
 So, the RG transformation is expressed in the form of some
 one-di\-men\-si\-o\-nal field theory with the ``hamiltonian'' (\ref{R}). The
 first line in Eq.(\ref{R}) represents the trivial diagonal part in the
 operation (\ref{RG}) while the second one represents the transposition
 between indices of ${\cal Q} $ in Eq.(\ref{RG}). The hamiltonian (\ref{R})
 differs only by the trivial diagonal part from that derived for representing
 the RG transformations for the composite operators of the unitary nonlinear
 $\sigma$ model that describes the quantum diffusion problem in the presence
 of a magnetic field.

 The operator (\ref{R}) which is just a representation of the simple
 symmetric group operation (all possible pair transpositions (\ref{RG}) in
 a set of $2s$ pairs of indices) can be diagonalized exactly
 \cite{AKL:91,Weg:90b}. The eigenvectors are given by
\begin{equation}
 \label{EV}
 |\vec\rho\,\rangle = \sum_{\{\vec s\,\}}g(\vec s\,)\chi_\rho(\vec s\,)\s
 \end{equation}
 where the summation is performed over all the partitions
 $\vec{s}\equiv\row{s}{m}\ldots$ of $s$ obeying the constraint (\ref{sm}),
 $
 g(\vec{s}\,)={s!}/{\prod_{m}m^{s_m}s_m!}
 $
 \ is the number of elements in the class defined by the partition $\vec s$,
 and $\chi_\rho(\vec s)$ are the characters of irreducible representation of
 the symmetric group characterized by the Young frame $\vec\rho$ having boxes
 of length $\row\rho m$ where $\sum_{m}\rho_m=s$. The appropriate eigenvalues
 are given by
 \begin{equation}
 \label{EgV}
 \alpha_s(\vec\rho\,)=\frac{s(s-1)}{2}+\sum_{m}\frac{\rho_m(\rho_m-2m+1)}{2}%
 \end{equation}
 To verify Eqs.(\ref{EV}) and (\ref{EgV}), it is necessary to remember that
 the operator (\ref{R}) is non-Hermitian, and $a^{\dagger}$ and $a$ defined
 by Eq.(\ref{aa}) are not mutually conjugate. In view of this, the set of
 bra-vectors orthogonal to the ket-vectors (\ref{EV}) is given by
 \begin{eqnarray}
 \langle\vec\rho|=(1/s!)\sum_{\{\vec s\,\}}\chi_\rho(\vec s\,)\langle\vec s|.
 \label{bra}
 \end{eqnarray}

 The maximum eigenvalue corresponds to the eigenvector characterized by the
 one-line Young frame with $\rho_1=s$, $\rho_m=0$ for $m>1$ for which $\chi
 _\rho (\vec s)=1$ for all $\vec s$, so that it is equal to $s(s-1)$, as in
 the case of the quantum diffusion. Note that it could be verified without
 any reference to the representations of the permutation group, acting by the
 transverse operator $\R$ on the bra-vector (\ref{bra}) which is proportional
 in this case just to $\sum \langle\vec s|.$ Thus, with the one-loop
 accuracy, the dimension of the operators coupled to the moments of the
 diffusion coefficient is given by
 \begin{equation}
 \label{AD}
 \alpha_s=-(s-1)d +\G_0s(s-1),%
 \end{equation}
 very similar to the case of the quantum diffusion, Eq.(\ref{ND}).

 Note that the minimum eigenvalue in Eq.(\ref{EgV}) (which corresponds to the
 eigenvector characterized by the one-column Young frame) equals zero. As
 results from Eq.(\ref{RG}), the eigenvalues for the solenoidal random-drifts
 differ in sign from those for the potential case. Thus, there is no positive
 exponents in the solenoidal case so that the dimensions of the operators
 coupled to the diffusion cumulants remain negative, and the distribution
 scales to the normal one. Therefore, of the three random-drifts model
 considered, the distribution becomes nontrivial and similar to the
 quantum-diffusion case only for the potential drifts.

 \section{ Comparison of the results for the quantum and classical diffusion}

 There are two types of contributions into the fluctuations of the diffusion
 coefficient in the random-walks model considered, similar to the quantum
 diffusion problem, Eq.(\ref{gn}). The ``normal'' one is given only by the
 functional (\ref{Si}). It diverges in the infrared limit thus making the
 fractional fluctuation $\cum{(\delta D)^2}/D^2\propto \G_0^2$ to be
 independent of the size of the system, analogous to the universal
 conductance fluctuation (the constant $\G_0$ has nonuniversal dependence on
 the disorder parameter, Eq.(\ref{g}), which restricts the analogy). This
 contribution has been considered in detail in ref.\cite{KLY:85} and is of no
 interest for the present considerations. The additional contribution to
 the diffusion cumulants (\ref{DCu}) is governed by the dimensions of the
 couplings $g^{s}$ in the ``additional'' functional (\ref{Sa}). Keeping only
 the maximum eigenvalue, as in Eq.(\ref{AD}), one finds in the critical
 dimensionality $d=2$ %
 \begin{equation}
 \label{Dc}
 \Cum{\lr(){\delta D}^s}\propto \lr(){\frac{l}{L}}^{2(s-1)-\GG_0s(s-1)}
 \end{equation}
 It is quite similar to the ``additional'' contribution to the conductance
 cumulants, Eq.(\ref{gn}b), in the quantum-diffusion problem (as the
 appropriate contribution to the cumulants of the density of states has the
 same form \cite{AKL:91}, one can write down the same expressions directly
 for the quantum-diffusion cumulants). To make the analogy more striking, one
 substitutes into Eq.(\ref{gn}b) the value of the parameter $u$,
 Eq.(\ref{u}), in the weak-localization limit at $d=2$,
 $u={g}_0^{-1}\ln(L/l)\equiv \widetilde u$, which gives
 \begin{eqnarray}
 \label{gnb}
 \Cum{g^s}\propto \lr(){\frac{l}{L}}^{2(s-1) - \overline{g}_0^{-1}s(s-1)}.
 \end{eqnarray}
 Note that both expressions (\ref{Dc}) and (\ref{gnb}) are applicable only
 for the high-order cumulants as all the eigenvalues of the RG equations but
 the maximum ones have been neglected in the derivation. Nevertheless, even
 for small $s\geq 2$ these expressions give a qualitatively valid result that
 the low-order cumulants are small as compared to the variance so that the
 bulk of the distribution is Gaussian. The high moments (with $s\gsim
 \G_0^{-1}$ for Eq.(\ref{Dc}) and $s\gsim \overline{g}_0$ in Eq.(\ref{gnb}))
 govern, in accordance with the identity (\ref{LI}), the lognormal tails of
 the distributions which have the following shape
 \begin{eqnarray}
 \label{DF}
 f(\delta X)\propto \frac{\alpha^2}{\delta X}\exp\lr[]{-\frac{1}{4\widetilde
 u}\ln^2\lr(){\delta X\alpha^{2} e^{-\widetilde u}}} %
 \end{eqnarray}
 An appropriate derivation is described in detail in ref.\cite{AKL:91}. Here
 $\delta X$ stands for either $\delta D$ or for $g$ in the classical or
 quantum problems, respectively, $\alpha\equiv L/l$, and $\widetilde u $ is
 given above for the quantum problem, and equals $\G_0\ln(L/l) $
 for the classical one.

 In the classical case, $\widetilde u$ has only a trivial dependence on the
 scale given by the logarithm, as there is no renormalization of the coupling
 constant. In the quantum case, $\widetilde u$ gives just the weak-disorder
 limit for the parameter $u=\ln(\sigma_0/\sigma)$, Eq.(\ref{u}). So the whole
 dependence on the renormalization of the coupling constant is absorbed by
 the parameter $u$ which is substituted to the lognormal distribution, its
 appearance being governed by the renormalization of the high-gradient
 operators. That is why one can separate the renormalization of the average
 values from that of the higher order cumulants. In such a way, one
 reproduces\cite{AKL:91} exact one-dimensional results for lognormal
 distributions by substituting the exact one-dimensional value of $u$ into
 the formulae similar to (\ref{DF}). It provides a basis for the conjecture
 that one can obtained a reasonable description of the distributions near the
 transition just by substituting $\exp(u)\sim |g-g_c|^{\nu}$ with a proper
 choice of the critical exponent $\nu$.

 \section{ Discussion}
 It has been demonstrated that the shape of the distribution of the diffusion
 coefficient in the model of classical diffusion in a medium with quenched
 random drifts proved to be very similar for that of the conductance
 distribution in a weakly disordered metal. In both cases the distributions
 turn out to be almost Gaussian in the weak-disorder limit but have slowly
 decreasing lognormal tails, and the part of the tails increase with
 increasing the disorder. This similarity occurs although the coupling
 constants in the field-theoretical models describing the quantum and the
 classical diffusion behave very differently. The asymptotic freedom of the
 nonlinear $\sigma$ model, \ie the increase of the coupling constant
 (inversely proportional to the conductance) with increasing a scale, is
 believed to govern the Anderson transition. No transition occurs in the
 classical diffusion problem described by the field theory with the
 $(\partial \phi\,\phi)^2$-type interaction where in the case of the
 potential disorder a perturbative renormalization of the coupling constant
 proves to be absent thus leading to the sub-diffusion, Eq.(\ref{As}a). The
 reason for the similarity is that the RG equations governing the cumulants
 of the distributions in both cases are classified according to the same
 irreducible representations of the group of permutations. But deriving the
 RG equations for the high-gradient operators proved to be much easier in the
 classical-diffusion model. Then the higher-loop RG analysis of the
 high-gradient operators which could check a viability of the one-loop
 results seems hardly possible in the quantum model but could appear quite
 straightforward for the classical one. Thus studying the classical-diffusion
 problem described here gives a possibility to learn more about the quantum
 diffusion in disordered media.

\end{document}